# Deposition and photoluminescence of zinc gallium oxide thin films with varied stoichiometry made by reactive magnetron co-sputtering

Martins Zubkins[1,*], Edvards Strods[1], Viktors Vibornijs[1], Anatolijs Sarakovskis[1], Ramūnas Nedzinskas[1], Reinis Ignatans[1], Edgars Butanovs[1], Juris Purans[1], Andris Azens[2]

[1]Institute of Solid State Physics, University of Latvia, Kengaraga 8, LV-1063, Riga, Latvia
[2]AGL Technologies SIA, Smerla 3, LV-1006, Riga, Latvia
*Corresponding author: martins.zubkins@cfi.lu.lv

**Abstract**

This paper reports on the deposition and photoluminescence of amorphous and crystalline thin films of zinc gallium oxide with Ga:Zn atomic ratio varied between 0.3 and 5.7. The films are prepared by reactive direct current magnetron co-sputtering from liquid/solid gallium/zinc targets onto fused quartz substrates; the temperature of the substrate is varied from room temperature (RT) to 800°C. The sputtering process is effectively controlled by fixing the sputtering power of one of the targets and controlling the power of the other target by plasma optical emission spectroscopy. The method, in conjunction with oxygen flow adjustment, enables the production of near-stoichiometric films at any temperature used. The composition analysis suggests a few at.% oxygen deficiency in the films. The resulting deposition rate is at least an order of magnitude higher compared to the commonly used radio-frequency sputtering from a ceramic $ZnO:Ga_2O_3$ target. Deposited onto unheated substrates, the films with Ga:Zn ≈ 2 are X-ray amorphous. Well-defined X-ray diffraction peaks of spinel $ZnGa_2O_4$ start to appear at a substrate temperature of 300°C. The surface of the as-deposited films is dense and exhibits a fine-featured structure observed in electron microscopy images. Increasing the deposition temperature from RT to 800°C eliminates defects and improves crystallinity, which for the films with Ga:Zn ratio close to 2 results in an increase in the optical band gap from 4.6 eV to 5.1 eV. Room temperature photoluminescence established the main peak at 3.1 eV (400 nm); a similar peak in $Ga_2O_3$ is ascribed to oxygen-vacancy related transitions. A prominent feature around 2.9 eV (428 nm) is attributed to self-activation center of the octahedral Ga-O groups in the spinel lattice of $ZnGa_2O_4$. It was found that photoluminescence from $ZnGa_2O_4$ depends significantly on the stoichiometric ratio between Ga and Zn and the deposition/annealing temperature.

# 1. Introduction

Ultra-wide band gap (UWBG) oxide semiconductors offer a wide range of possible applications and are currently under extensive research [1,2]. The applications cover the fields of solar-blind photodetectors, high-voltage transistors, gas sensors, and light-emitting diodes. Zinc gallate ($ZnGa_2O_4$) is a complex oxide compound with an optical band gap in the range of 4.4–5.0 eV [3]. Additional properties such as high electron mobility of ≈100 $cm^2/V·s$ [4] and thermal conductivity of 22 W/m·K [5] as well as high chemical and thermal stability make it a good candidate for the UV optoelectronic devices. Furthermore, p-type conductivity has recently been demonstrated in $ZnGa_2O_4$ thin films [6,7], enabling their applicability in bipolar power electronics devices [8]. It has a spinel structure (space group *Fd*3*m*, *a*=8.335 Å) where Zn and Ga cations occupy tetrahedral and octahedral sites, respectively [9].

$ZnGa_2O_4$ is a blue-emitting phosphor [10], and a significant energy difference between the absorption and emission features, known as the Stokes shift, is observed in $ZnGa_2O_4$ films [11]. A large Stokes shift reflects the typical property of Ga-based UWBG semiconductors and can be associated with self-trapped holes formed in the deep energy levels. It is likely related to the difficulties in achieving technologically desirable p-type doping [12]. For Ga-based UWBG oxide semiconductors, the lower the Ga(III) coordination number, the higher the energy of the first absorption band, resulting in a larger Stokes shift [13]. Photo- and cathodoluminescence studies of $ZnGa_2O_4$ films doped with transition and rare-earth metal ions have shown efficient luminescence of the material making it suitable for a variety of optical applications [14–17].

Thin films of $ZnGa_2O_4$ can be grown by several different deposition techniques, including metal-organic chemical vapor deposition [18], pulsed laser deposition (PLD) [19], atomic layer deposition [20], and radio-frequency (rf) magnetron sputtering [21]. Magnetron sputtering typically provides high deposition rates, well-controlled and reproducible conditions, low processing temperatures, dense and smooth surfaces, and scalability [22]. The magnetron sputtered films in most cases are amorphous or polycrystalline. High crystallinity of $ZnGa_2O_4$ films can be obtained by rapid thermal annealing [23]. However, only rf sputtering from ceramic targets has been used to grow $ZnGa_2O_4$ films so far. This type of sputtering has significant drawbacks to other forms of sputtering, including lower deposition rates and power efficiency, and higher costs [24]. Recently, our group has demonstrated a high-rate deposition technique of stoichiometric $Ga_2O_3$ amorphous, polycrystalline and epitaxial thin films using reactive pulsed direct current (pulsed-DC) magnetron sputtering from a liquid Ga metal target [25]. The method also enables the growth of Ga containing complex oxide films by co-sputtering in a reactive atmosphere, as demonstrated in the case of a $Ga_2O_3$-$ZnGa_2O_4$ core-shell nanowire heterostructure production [26].

This paper reports on the growth of zinc gallium oxide thin films in Ga:Zn concentration ratio range between 0.3 and 5.7 on fused quartz substrates. The films are deposited by reactive DC magnetron co-sputtering from a-zinc and a liquid gallium metal target in an argon and oxygen atmosphere. The effects of the sputtering power, oxygen gas flow, and substrate temperature on the film composition, structure, optical properties, and photoluminescence are investigated.

## 2. Experimental details

Film deposition was performed using a G500M.2 physical vapor deposition coater equipped with two planar balanced magnetrons (Sidrabe Vacuum, Ltd.) with target dimensions 150 mm × 75 mm. The films were deposited by reactive DC magnetron co-sputtering from a liquid gallium (purity 99.999%) and a solid zinc (purity 99.95%) metal targets in an Ar/O$_2$ atmosphere. A schematic of the deposition system is shown in Fig. 1(a). The preparation of the gallium target has been described in detail elsewhere [25]. The magnetrons were placed against the grounded substrate holder at a distance of approximately 11 cm. The base pressure was ≤7×10$^{-4}$ Pa in a turbomolecular pumped chamber. The process pressure was set to 0.47 Pa by feeding 30 sccm (standard cubic centimeters per minute) of argon (99.9999% pure) and 11–15 sccm of oxygen (99.999% pure), and partly closing the throttle valve between the chamber and the pump. Fused (f-) quartz (SPI Supplies) substrates were used. The substrate temperature was varied between room temperature, i.e. without intentional heating, and 800°C. A few selected samples deposited at room temperature (RT) were post-annealed for 3 h at 700°C in air, to determine whether there were differences in the properties of films heated during or after the deposition.

The sputtering process was controlled by optical emission spectroscopy (OES). The plasma OES from both discharges, i.e., Ga and Zn, were collected separately (shown in Fig. S1 in the supplementary material) by a multichannel PLASUS EMICON MC spectrometer (spectral range of 200–1100 nm). Optical fiber probes were placed in the chamber overlooking the discharges parallel to the target surface 1.5 cm above the target erosion zone. To vary the composition of the films, the gallium 417.2 nm and zinc 481.1 nm emission line intensity ratio $I_{Ga417}/I_{Zn481}$ was changed from sample to sample by a variation of the sputtering power of Zn ($P_{Zn}$) upon the sputtering power of Ga fixed at 100 W (0.89 W/cm$^2$). During a single deposition cycle, $I_{Ga417}/I_{Zn481}$ was kept constant by fine tuning of $P_{Zn}$. An example of a typical deposition cycle is shown in Fig. 1(b).

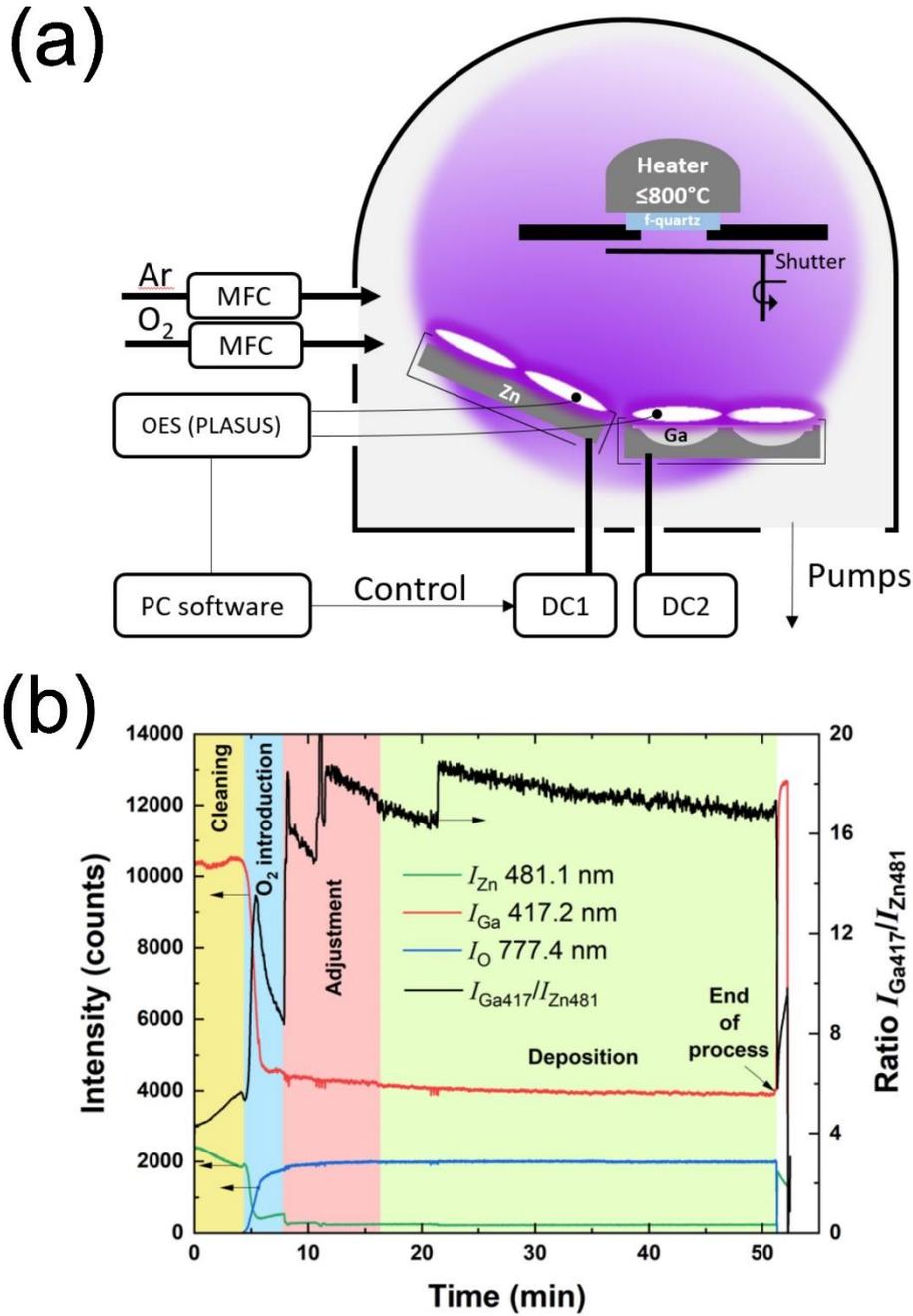

**Fig. 1.** (a) Schematic of the co-sputtering system. (b) Example of a single deposition cycle controlled by optical emission spectroscopy. The process was started in the metal mode (feeding argon only) with the shutter closed in front of the substrate. In the beginning, $P_{Zn}$ was 100 W (0.89 W/cm$^2$). The targets were cleaned for approximately 5 min. Oxygen was gradually added to the selected value and kept constant thereafter. Then, $P_{Zn}$ was adjusted to reach the desired $I_{Ga417}/I_{Zn481}$ value of approximately 17. The shutter was opened, and the film was deposited onto the unheated (RT) substrate. $I_{Ga417}/I_{Zn481}$ was kept constant during the deposition by adjusting $P_{Zn}$. In this specific example, $P_{Zn}$ during the deposition was approximately 65 W (0.58 W/cm$^2$), and the deposition time was 35 min. The atomic ratio Ga:Zn in the film measured after the process was found to be equal to 2.

For comparison purposes, we deposited a zinc gallium oxide film (121 nm) by rf magnetron sputtering from a ceramic two-inch ZnO:Ga$_2$O$_3$ (1:1, purity 99.99%) target in an Ar/O$_2$ atmosphere (30/10 sccm, purity 99.9999%/99.999%) on an f-quartz substrate at 350°C. The deposition process was carried out using the R&D vacuum coater SAF25/50 (Sidrabe Vacuum, Ltd.), equipped with a circular, balanced magnetron from Gencoa, Ltd. Sputtering was conducted at a power of 90 W (4.4 W/cm$^2$), and the pressure was set to 0.47 Pa by partially closing the throttle valve. The distance between the target and the substrate was approximately 13 cm.

X-ray photoelectron spectroscopy (XPS) was used to study the chemical composition of the samples. The XPS measurements were carried out using a ThermoFisher ESCALAB Xi+ instrument using a monochromatic Al Kα X-ray source. The charge neutralization by flow of low energy electrons was used in the reported experiments. The spectra were recorded using an X-ray beam size of 650 × 100 μm with a pass energy of 20 eV and a step size of 0.1 eV.

The crystallographic structure of the films was examined by X-ray diffraction (XRD), using a Rigaku MiniFlex600 X-ray powder diffractometer with Bragg-Brentano θ-2θ geometry and a 600 W Cu anode (Cu Kα radiation, λ = 1.5406 Å) X-ray tube.

The film thickness and optical properties were established by WOOLLAM RC2 spectroscopic ellipsometer (SE) in the range from 225 to 827 nm (5.5–1.5 eV). The main ellipsometric angles Ψ and Δ were measured at the incident angles from 55° to 65° with a 5° step. Refractive index $n$ and extinction coefficient $k$ dispersion curves were modelled using Cauchy and B-Spline Kramers–Kronig consistent B-spline models. SE experimental data model-based regression analyses were performed with the WOOLLAM software CompleteEASE®. The mean squared error (MSE) values of the models ranged from 1 to 10. The thicknesses of the studied films are presented in Tables S1 and S2, with an average value of 260 nm and a standard deviation of 66 nm (26%) attributed to variations in deposition conditions and time.

The room-temperature photoluminescence (PL) measurements were recorded using a custom-made setup. PL was excited using a pulsed solid-state 266 nm laser (4-th harmonic Nd:YAG; > 0.5 μJ). The laser beam was unfocused to avoid damaging the sample. The power density was controlled between 90-650 kW/cm$^2$ with fused-silica gradient filter. PL was dispersed by using the 500 mm focal length monochromator (Andor SR-500i; UV-VIS gratings: 1200 g/mm blazing at 300 nm; 600 g/mm blazing at 1000 nm) and focused into photomultiplier tube (PMT) (ACC-SR-ASM-0047; Oxford instruments). The conventional lock-in detection system (SR830; Stanford Research Systems) was used to extract the emission signal. Further experimental details can be found in Ref. [27].

## 3. Results and discussion

*3.1 Influence of deposition conditions on chemical composition*

First, the $I_{Ga417}/I_{Zn481}$ ratio and the substrate temperature were varied to determine the influence of the deposition conditions on the film composition. The Ga:Zn atomic ratio measured by XPS as a function of $I_{Ga417}/I_{Zn481}$ and substrate temperature at a constant oxygen flow of 11 sccm is shown in Fig. 2(a), where larger values of $I_{Ga417}/I_{Zn481}$ correspond to lower $P_{Zn}$. It is evident that the composition can be efficiently controlled by $P_{Zn}$. In the case of substrate temperature of RT and 600°C, the increase of Ga:Zn atomic ratio within the studied range of process parameters is linear with the $I_{Ga417}/I_{Zn481}$ ratio. This is understandable because the amount of emitted light detected by OES is, within a reasonably accurate approximation (potentially ignoring some higher order effects such as the optical excitation yield changing with the sputtering power or the degree of the target surface oxidation), directly proportional to the number of sputtered atoms and hence the number of atoms reaching the substrate. The results are consistent with the linear relationship seen between the OES signal and the growth rate of gallium oxide films at RT [25].

The film composition depends on the substrate temperature as well. Increasing the temperature from RT to 800°C and keeping $I_{Ga417}/I_{Zn481}$ in the range of 2.5 to 3.0, the Ga:Zn atomic ratio increases from 0.6 to 4.9 as indicated by the vertical arrow in Fig. 2(a). Since the sputtering conditions do not change, the arrival rate of atoms at the substrate should be the same at any temperature. Apart from Zn atoms sticking to the substrate less than Ga atoms because of the higher vapor pressure of Zn compared to that of Ga [28] (before adatoms are oxidised), our results suggest that the difference in sticking probability is temperature dependent, and, if not compensated by other deposition parameters, this leads to the change in stoichiometry with the substrate temperature. The results are consistent with Zn deficiency observed in $ZnGa_2O_4$ films deposited from stoichiometric $ZnO:Ga_2O_3$ targets [19,23,29,30]. We note that the possibility of counteracting the effect of the substrate temperature by adjustment of $P_{Zn}$, is a major advantage of co-sputtering compared to the single target sputtering, which would require different target compositions for obtaining the same film composition at different substrate temperatures.

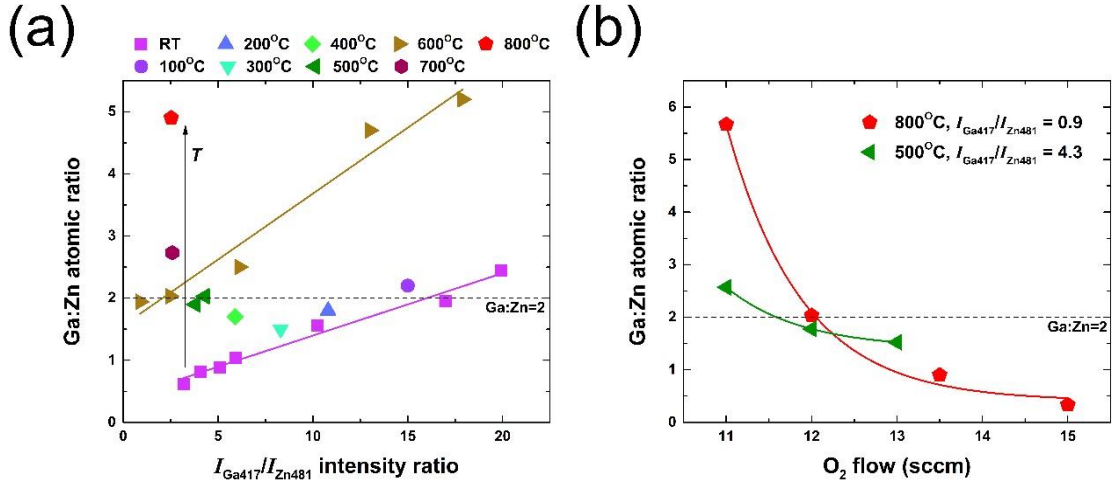

**Fig. 2.** (a) Ga:Zn atomic ratio as a function of $I_{Ga417}/I_{Zn481}$ for the samples deposited at different substrate temperatures while keeping the oxygen flow rate at 11 sccm. The vertical black arrow shows the substrate temperature influence on Ga:Zn when $I_{Ga417}/I_{Zn481}$ is fixed in the range 2.5–3.0. (b) Ga:Zn as a function of the oxygen flow at two deposition conditions – 500°C ($I_{Ga417}/I_{Zn481}$=4.3) and 800°C ($I_{Ga417}/I_{Zn481}$=0.9).

It is evident from Fig. 2(a) that the dependence of film composition on the substrate temperature is strongly non-linear. The data points recorded at 100°C, 200°C and 300°C do not deviate significantly from the RT trendline, suggesting that there is no major influence of the temperature on the composition in this temperature interval. The points at 400°C to 700°C are placed increasingly farther from the RT trendline with roughly similar increments per 100°C change in temperature, followed by a significantly larger increase in Ga:Zn ratio upon going from 700°C to 800°C substrate temperature. This observation, also depicted in Fig. S2, aligns with the trend presented in Ref. [31], where the sticking coefficient is plotted against substrate temperature. The function can be divided into three regions: (i) the complete adsorption region, (ii) the no-adsorption region, and (iii) the sharp transition region, where the sticking coefficient undergoes a significant reduction from near unity to almost zero with increasing temperature. Moreover, the stoichiometry and morphology of the growth interface can influence the sticking coefficient as deposition progresses, given its impact on binding energy [32,33]. This effect can lead to a non-constant sticking coefficient for weakly bonding atoms over time. For this reason, the composition depth profile should be a matter of further study. At low temperatures, the stoichiometry-dependent desorption becomes negligible.

The 3-hour post-annealing at 700°C in an oxidizing atmosphere (air) had no measurable effect on the composition of the films. The different impact of the temperature upon deposition and annealing is thought to arise from the fact that during the annealing of an already oxidised film Zn and Ga atoms are bonded with oxygen. During the deposition these atoms adsorb onto the substrate surface and may desorb again before reacting with oxygen.

We also note that there are practical limits to how far the Ga:Zn ratio in the films can be varied just by varying the sputtering power to the zinc target, $P_{Zn}$. For example, to deposit a film with Ga:Zn ratio of 2 at the substrate temperatures of 700°C and 800°C, it would be necessary to reduce $I_{Ga417}/I_{Zn481}$, i.e., to increase $P_{Zn}$ to a level exceeding the limit of safe operation. A common upper limit for DC magnetron sputtering, before target damage sets in, is approximately 0.05 kW/cm$^2$ [34]; however, this limit strongly depends on factors such as the target material and quality, the type of target cooling, and the construction of the magnetron itself. The problem can be avoided by using different oxygen flows at different substrate temperatures. Fig. 2(b) depicts the Ga:Zn atomic ratio at different oxygen flows for two different deposition conditions – 500°C ($I_{Ga417}/I_{Zn481}$=4.3) and 800°C ($I_{Ga417}/I_{Zn481}$=0.9). Zn volatilization is reduced at higher oxygen flows by more efficient oxidation on the substrate surface. This is particularly evident at 800°C, where Ga:Zn decreases from 5.7 to 0.3 when the oxygen flow is increased from 11 to 15 sccm. At 500°C, it is less pronounced but the effect is still observable.

The composition of the samples deposited at RT, 600°C, and 800°C is plotted in the Zn-Ga-O chemical composition diagram (Fig. 3), which shows the ability of the reactive co-sputtering to produce zinc gallium oxide films in a wide range of composition. However, in the films with Ga:Zn close to 2, the XPS measurements suggest a relatively large 2–5 at.% oxygen deficiency.

An additional advantage of the reactive co-sputtering method over rf sputtering from a ceramic target is the significantly higher sputtering/deposition rate. The static deposition rate varies from about 5 to 14 nm/min depending on $P_{Zn}$, thus $I_{Ga417}/I_{Zn481}$ ratio and also substrate temperature, as shown in Fig. S3. The rate increases with increasing $P_{Zn}$ or decreasing temperature. The deposition rate for the rf-sputtered film was determined to be approximately 0.4 nm/min. This rate was found to be at least an order of magnitude lower than that of reactive co-sputtering, despite the sputtering power density being roughly two times larger when compared to the co-sputtering (summing both power densities on the Zn and Ga targets together) with just a two cm larger distance between the target and the substrate. Additionally, the rf-sputtered film was measured to be highly non-stoichiometric, with the Ga:Zn ratio of approximately 4. A somewhat higher static deposition rate of 1.4 nm/min on c-sapphire and silicon (100) wafer substrates in the temperature range of 200–600°C is reported in Ref. [21] for rf magnetron sputtering from a ceramic ZnGa$_2$O$_4$ target (ZnO:Ga$_2$O$_3$ mixed powder, 30:70 proportion). The higher deposition rate can be partly explained by the sputtering power of 7.4 W/cm$^2$, which is 1.7 times higher than in our case. However, the deposition rate is still significantly lower compared to reactive DC co-sputtering.

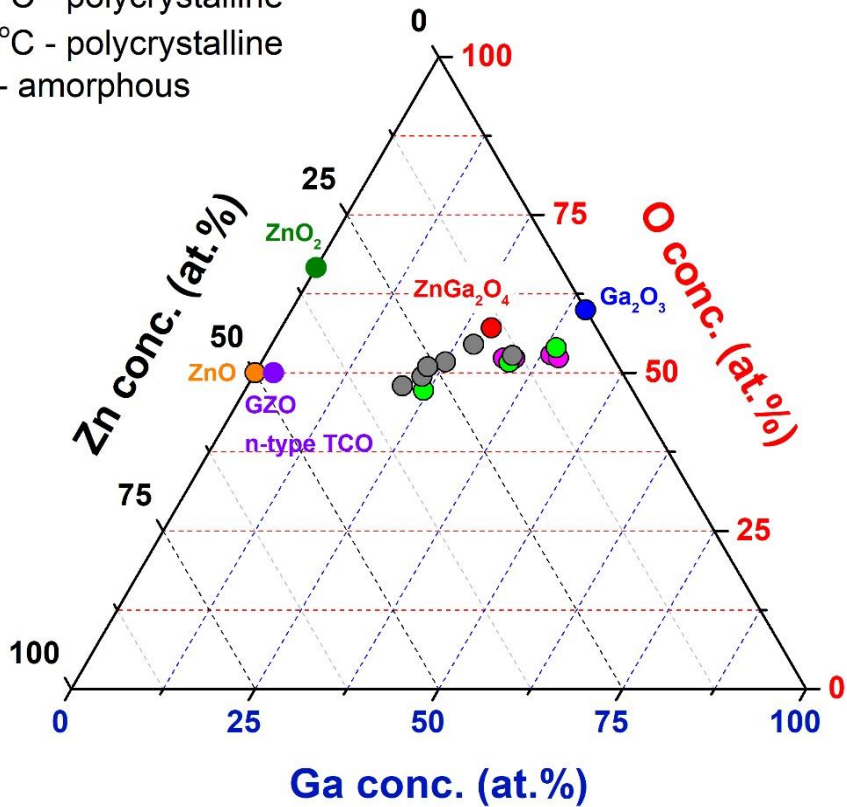

**Fig. 3.** Composition measured by XPS of the films deposited at RT (grey circles), 600°C (pink circles), and 800°C (green circles) and represented in a ternary diagram of Zn, Ga, and O. Reference compositions – stoichiometric $ZnGa_2O_4$ (red circle), $Ga_2O_3$ (blue circle), ZnO (orange circle), $ZnO_2$ [35] (dark green circle), and transparent n-type conducting Ga doped ZnO (GZO [36], violet circle).

*3.2 Influence of substrate temperature and chemical composition on structure*

Initially the deposition process was developed by producing the films at RT. The XRD patterns (recorded over a 2θ-range of 5°–90°) of these films as a function of Ga:Zn is shown in Fig. S4(a). The XRD patterns for the films with Ga:Zn ≥ 1 do not show Bragg peaks associated with any of the possible oxide phases suggesting that the films are X-ray amorphous. A wide XRD peak around 32° appears when Ga:Zn is in the range of 0.6–1.0. It corresponds to the (100) plane of the wurtzite ZnO structure according to ICDD card 01-070-8072 and indicates nanocrystalline phase formation with crystallite size of 4.5 nm calculated from the Scherrer equation. In addition, the intensity of the peak is lower than the intensity of the halo peak around 22° attributed to the substrate, which is indicative of the predominantly amorphous phase in the films. Fig. S4(b) shows the XRD patterns of the post-annealed films deposited at RT with two different Ga:Zn ratios of 0.7 and 2.0. The atoms in the amorphous structure of the films rearranged during post-annealing and formed crystalline phases. In the case of Ga:Zn=0.7, a mixture of crystalline spinel $ZnGa_2O_4$ and wurtzite ZnO phases was observed. In the case of Ga:Zn=2.0, we observed only the formation of crystalline $ZnGa_2O_4$. It has been claimed that a long annealing time can lead to phase separation and defect formation due to the diffusion effect [23]. Therefore, the formation of either $Ga_2O_3$ or ZnO nanocrystallites cannot be ruled out, as they would not be detectable by XRD.

Fig. 4(a) shows XRD patterns of the films deposited in the substrate temperature range RT–800°C while keeping Ga:Zn close to 2. Crystallization is detectable starting at 200°C and improves at higher temperatures. The XRD patterns reveal the formation of a crystalline spinel $ZnGa_2O_4$ phase. All the observed XRD peaks at 18.6°, 35.9°, 37.5°, 43.3°, and 57.4° for the films deposited between 300°C and 800°C correspond to cubic $ZnGa_2O_4$ (111), (311), (222), (400), and (511) planes, respectively, according to the powder diffraction ICDD card 01-071-0843. The XRD pattern of the rf-sputtered film, showing the crystalline $ZnGa_2O_4$ phase, is presented in Fig. S5(a). As the substrate temperature increases from 500°C to 800°C, the texture of the co-sputtered films changes slightly, which is evident by comparing the intensity ratios between different peaks. It can be argued that the peak at 18.6° and the peak with low intensity at 30.6° might also correspond to the β-$Ga_2O_3$ (-201) and (-401) planes at 18.95° and 30.51° according to ICDD card 01-087-1901. The angle of the more intense peak gradually shifts closer to 18.95° ((-201) of β-$Ga_2O_3$) from 18.3° to 18.7° as the temperature increases from 200°C to 800°C. Indeed, the formation of a crystalline $Ga_2O_3$ phase in the $ZnGa_2O_4$ films grown at substrate temperatures above 700°C has been proven and explained by the loss of Zn [23,29]. However, in our case, we compensate for this by increasing $P_{Zn}$ and oxygen flow. Additionally, the intensity of these XRD peaks does not correlate with the Ga:Zn ratio as might be expected. Fig. 4(b) depicts XRD patterns of the films deposited at 600°C as a function of Ga:Zn. The peak at 18.6° is most intense at Ga:Zn of 2.0 and vanishes completely at Ga:Zn of 5.2. Fig. 4(b) also shows that the XRD peaks of $ZnGa_2O_4$ are almost nonexistent and a wide peak with low intensity of (100) plane of the wurtzite ZnO structure appears at Ga:Zn of 1.3.

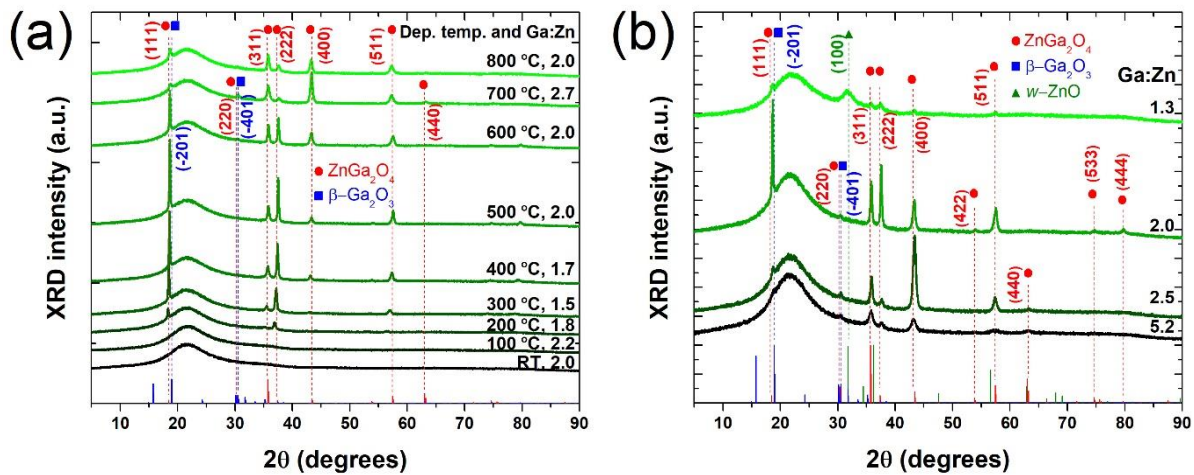

**Fig. 4.** XRD patterns of the zinc gallium oxide films deposited on f-quartz substrates as a function of (a) substrate temperature with Ga:Zn in the range of 1.5–2.7 and (b) Ga:Zn ratio for the films deposited at 600°C. The angles of the Bragg peaks for cubic spinel $ZnGa_2O_4$ (ICDD card 01-071-0843), monoclinic $\beta$-$Ga_2O_3$ (ICDD card 01-087-1901), and hexagonal wurtzite ZnO (ICDD card 01-070-8072) are shown by vertical lines with red circles, blue squares, and green tringles as indicators, respectively.

*3.3 Influence of substrate temperature and chemical composition on surface morphology*

Surface morphology SEM images of the zinc gallium oxide films at different deposition temperatures and different Ga:Zn ratios are shown in Fig. 5. The images of the films deposited at 200°C and 400°C (Fig. 5(a,b)) show a heterogeneous surface with distinct areas. As the deposition temperature increases, the surface becomes homogeneous and smooth, with a fine-featured structure (Fig. 5(c-e)). The rf-sputtered film deposited at 350°C exhibits a similar surface morphology but with visibly higher roughness (Fig. S5(b)). Furthermore, the surface features become larger as higher temperatures promote the mobility of atoms and improve the film's crystallinity. At 800°C, the surface exhibits grainy features with sharp edges. These results are consistent with observations from XRD patterns and demonstrate the possibility of controlling the crystallinity of the films with the deposition temperature. Overall, the observed dense microstructure of the films is as expected since the magnetron sputtering technique is known to produce dense coatings [37].

The effect of the Ga:Zn ratio on the surface morphology is evident when comparing the samples deposited at 600°C with Ga:Zn=2.0 or 4.7 with the sample with Ga:Zn=1.3 in Fig. 5(d,f,g). At Ga:Zn=1.3, the surface has more pronounced features with random shapes, making it rougher. The sample contains a mixture of both nanocrystalline spinel $ZnGa_2O_4$ and wurtzite ZnO phases according to the XRD pattern. The surface becomes visibly smoother when the Ga:Zn is increased. Fig. 5(h) shows the surface of the post-annealed film. After annealing, the film is cracked, and there are round particles on the surface with a size of tens of nanometers.

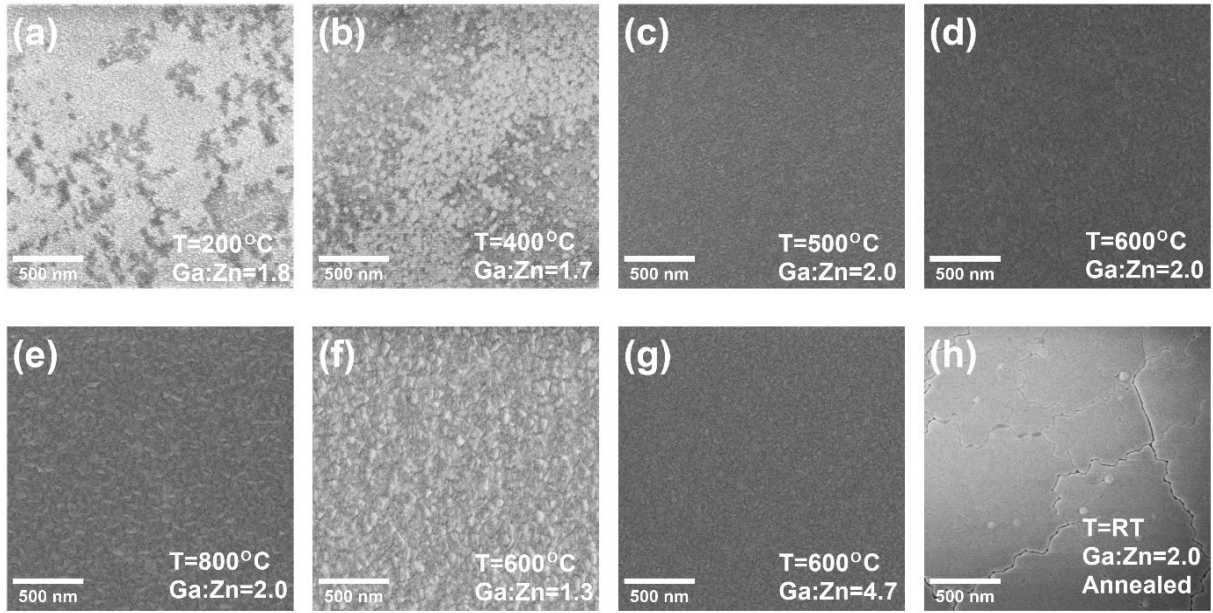

**Fig. 5.** Surface SEM images of the zinc gallium oxide films deposited by reactive magnetron co-sputtering on f-quartz substrates at different deposition temperatures – (a) 200°C, (b) 400°C, (c) 500°C, (d) 600°C, (e) 800°C, with Ga:Zn ratio in the range 1.7–2.0 and with different Ga:Zn ratios – (f) 1.3, (g) 4.7, deposited at 600°C. Image (h) depicts the surface of the post-annealed sample for 3 h in air at 700°C deposited at RT with Ga:Zn=2.

*3.4 Influence of substrate temperature and chemical composition on optical properties*

The optical properties of the zinc gallium oxide films were studied by SE. Refractive index $n$ and extinction coefficient $k$ dispersion curves as a function of photon energy $E$ (1.5–5.5eV) are given in Fig. 6. The optical constants depend on the Ga:Zn ratio as well as the deposition temperature. There is a tendency for $n$ to decrease with Ga:Zn ratio (Fig. 6(c,f)), which is particularly noticeable for the RT deposition, where $n$ at 2.25 eV decreases from 1.95 to 1.85 as Ga:Zn increases from 0.8 up to 2.5. The decrease of $n$ is consistent with the trend found in the literature, where the values of $n$ at 2.25 eV for wurtzite ZnO, spinel $ZnGa_2O_4$, and $\beta$-$Ga_2O_3$ are 2.00–2.10 [38,39], 1.90–1.95 [40,41], and 1.80–1.90 [42,43], respectively. Furthermore, $n$ at 2.25 eV increases rapidly from approximately 1.86 to 1.97 when the deposition temperature is increased from 100°C to 200°C (Fig. 6(i)). This temperature range coincides with the onset of crystallization of the $ZnGa_2O_4$ phase observed from XRD (Fig. 4(a)). The decrease is less pronounced at 600°C, where $n$ at 2.25 eV only decreases from 1.973 to 1.943 as Ga:Zn increases from 1.3 up to 5.2. The dependence of $n$ on the temperature above 200°C is not clear, as Fig. 6(i) shows that $n$ is more influenced by the chemical composition fluctuations – Ga:Zn in the range from 1.5 to 2.7.

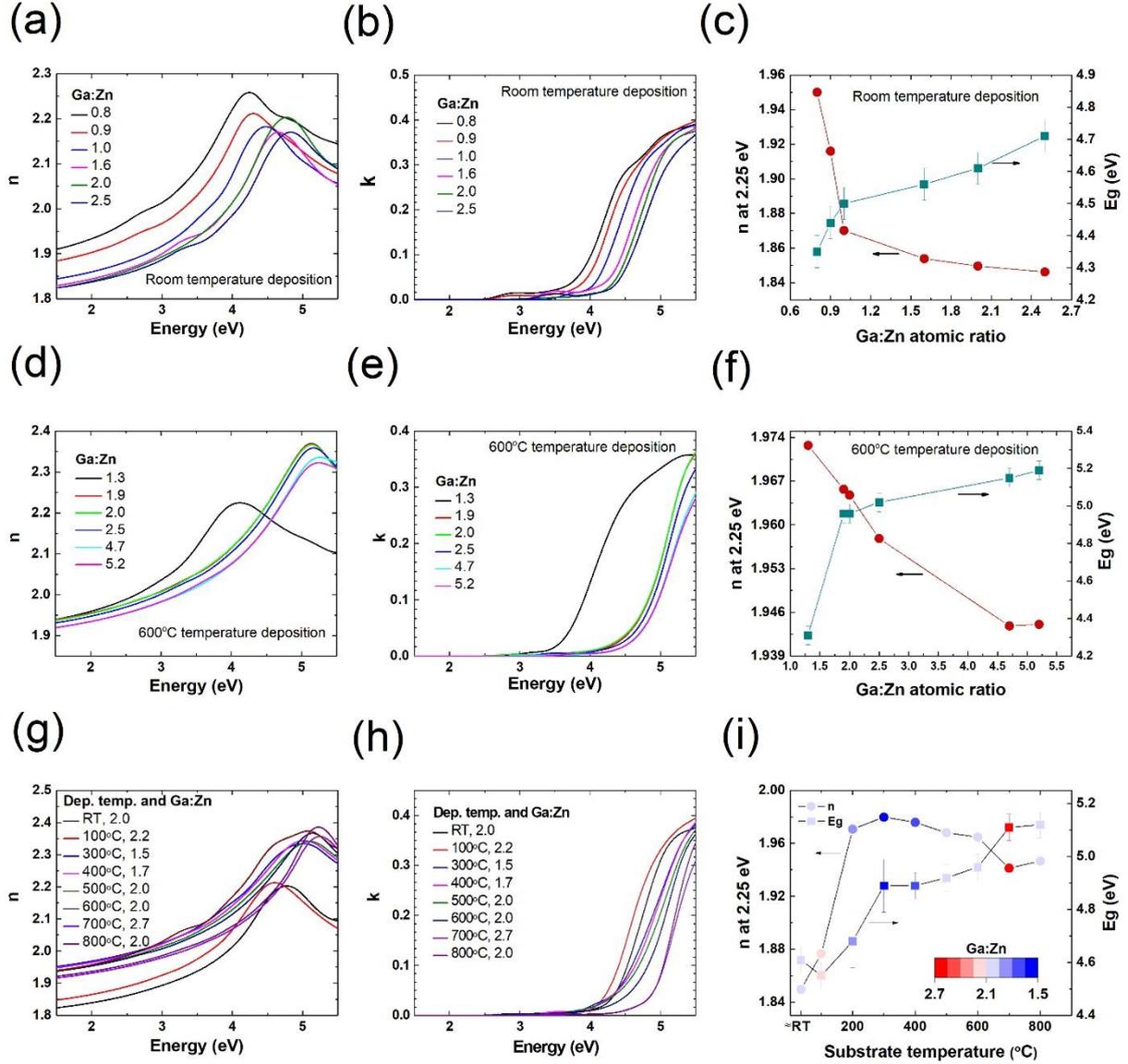

**Fig. 6.** (a,d,g) Refractive index $n$ and (b,e,h) extinction coefficient $k$ dispersion curves as a function of photon energy for zinc gallium oxide thin films deposited by reactive magnetron co-sputtering at different temperatures (RT–800°C) and with different Ga:Zn ratios (0.8–5.2). $n$ at 2.25 (550 nm) and optical band gap $E_g$ as a function of (c,f) Ga:Zn ratio (deposition at RT and 600°C, respectively) and (i) substrate temperature with Ga:Zn in the range 1.5–2.7.

The sharp increase in $k$ between 4 and 5 eV is due to the onset of the fundamental absorption of zinc gallium oxide. The optical energy gap ($E_g$) of the films was determined using the $k$ values and applying the Tauc plot ($\alpha h\nu)^2$ vs $h\nu$ for direct allowed transitions, where $h\nu$ represents the photon energy and $\alpha=4\pi k/\lambda$ represents the absorption coefficient. The $E_g$ of the films was estimated by extrapolating the linear part of ($\alpha h\nu)^2$ on $h\nu$ axis (Fig. S6). $E_g$ increases with both Ga:Zn ratio and deposition temperature (Fig. 6(c,f,i)). It shifts from 4.3 to 4.7 eV when Ga:Zn ratio is increased from 0.8 to 2.5 and from 4.3 to 5.2 eV when Ga:Zn ratio is increased from 1.3 to 5.2 for the films deposited at RT and 600°C, respectively. This again correlates with the trend of the $E_g$ values of 3.3–3.4 eV [38,39], 4.4–5.0 eV [3], and 4.9–5.1 eV [25,44] for pure wurtzite ZnO, spinel ZnGa$_2$O$_4$, and β-Ga$_2$O$_3$, respectively. The increase in $E_g$ from 4.6 to 5.1 eV as the deposition temperature increases from RT to 800°C while keeping Ga:Zn in the range of 1.5–2.7 is attributed to improved crystallinity with fewer defects and grain boundaries.

In addition to the fundamental absorption, the films deposited at RT have an additional absorption band in the visible (1.8–3.2 eV) or UVA (3.2–3.9 eV) light ranges, as shown in Fig. 6(b) and reported in Ref. [19]. The position of the absorption band depends on the Ga:Zn ratio. $k$ values of 0.010–0.015 peak at around 3 eV for the zinc rich films with Ga:Zn of 0.8 and 0.9. The band shifts into the UVA range when Ga:Zn ratio is increased. Increasing the deposition temperature above RT suppresses the band to $k$ values of 0.002–0.005 (Fig. 6(e,h).

Photoluminescence (PL) spectra for the zinc gallium oxide thin films deposited at different substrate temperatures between RT and 800°C are given in Fig. 7(a). All the emission spectra show a peak centered at approximately 3.1 eV, and a low photon energy tail pronounced especially for the samples grown at lowest temperatures (RT and 200°C). The peak at 3.1 eV is similar to the one in PL spectra from Ga$_2$O$_3$, typically dominated by broadened optical transitions centered near 3.11 eV (399 nm), which have been ascribed to oxygen-vacancy related transitions [45]. Note also the lowermost dotted spectrum in Fig. 7(a) showing the strong effect due to thermal annealing: the broad low photon energy tail for the film deposited at room temperature is significantly reduced by annealing at 700°C temperature, which correlates with crystallization of the initially amorphous film into the ZnGa$_2$O$_4$ phase, also in accordance to XRD data presented above. The previous report on the optical transmittance and emission blue-shift of randomly oriented ZnGa$_2$O$_4$ films, grown by rf sputtering, also indicates structural and optical improvements resulting from the conversion of their polycrystalline nature into a quasi-single-crystalline structure through thermal annealing treatment within the temperature range of 500-900°C [21]. Additionally, a similar enhancement of structural quality through high-temperature (800°C) thermal annealing was reported for single-crystalline ZnGa$_2$O$_4$ epitaxial thin films [46].

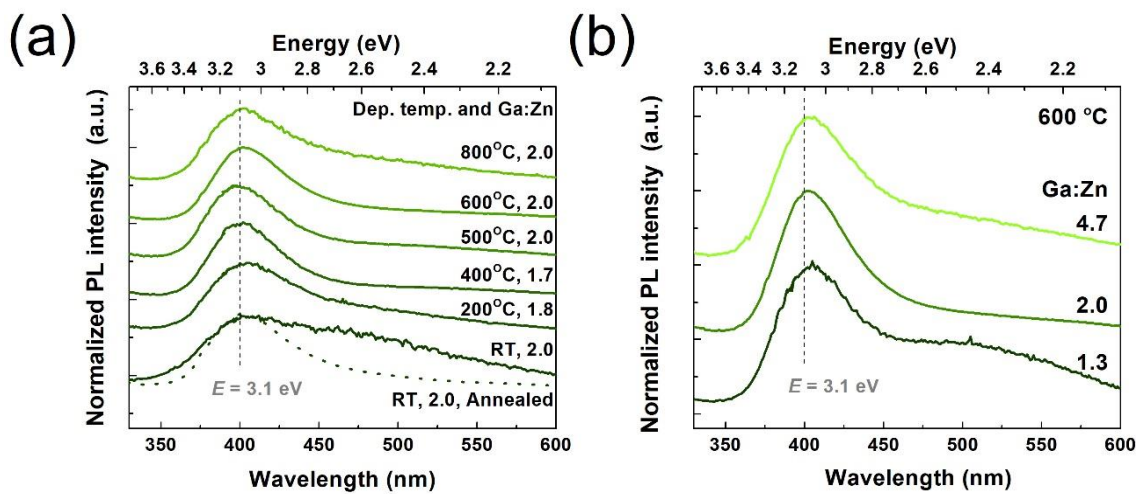

**Fig. 7.** Normalized room temperature (RT) PL spectra (a) for films of $ZnGa_2O_4$ grown at different substrate temperatures, and (b) films with different Ga:Zn ratios grown at 600°C temperature.

Fig. 7(b) shows PL spectra for films with different Ga:Zn ratios deposited at 600°C substrate temperature. Although the emission peak located at approximately 3.1 eV is not shifted with the composition, there is a low photon energy tail present in the spectra of the films with compositions different from stoichiometric $ZnGa_2O_4$. In particular, it can be noted that the "cleanest" spectrum with a single line centered at 3.1 eV is observed for nominally stoichiometric crystalline $ZnGa_2O_4$ films deposited at 400–600°C substrate temperature. Compositions other than stoichiometric and/or substrate temperatures outside the 400–600°C range yield spectra with the dominant peak at 3.1 eV accompanied by a broad tail stretching across the green part of the spectrum.

In order to get a deeper insight into the optical emission properties recorded for the nominally stoichiometric $ZnGa_2O_4$ thin films grown within temperature range of 500–800°C, Gaussian decomposition of the chosen PL spectra was carried out, as shown in Fig. 8. The narrowest and most intense PL band (Gauss-1) is located at 3.101 eV (399 nm), and its nature was briefly described earlier. A less intensive emission band (Gauss-2) towards lower energy range is centered at 2.9 eV, which is attributed to the emission in $ZnGa_2O_4$. In particular, by exploiting the absorption band around 260 nm [47], i.e., with the use of a 266 nm laser excitation source, the luminescence band around 2.870 eV (432 nm) in $ZnGa_2O_4$ is established, which is originated from self-activation center of the octahedral ($O_h$) Ga−O groups in the spinel lattice [48,49]. This assignment is further supported by an optical study of $ZnGa_2O_4$ phosphors, synthesized via the high-temperature solid-state reaction method, wherein the dominant emission band at 2.88 eV (430 nm) was recorded [50]. Alternatively, this faint optical feature, recorded very close to the energy of 2.981 eV (416 nm), can be ascribed less likely to electron-hole recombination in $Ga_2O_3$ formed by Ga-O vacancy pair (between the $V_O$ donor band and $V_{Ga}$ acceptor band), as suggested in Ref. [51,52]. It is also unlikely to be related to defect-induced optical transitions, as proposed in Ref. [21], where very faint PL features were recorded for rf magnetron-sputtered $ZnGa_2O_4$ films. Furthermore, the very wide and low intensity band (Gauss-3) at 2.50–2.64 eV is attributed to tunnel capture of an electron from a donor cluster by a hole on an acceptor [53]. Indeed, this broad feature was recorded to be growing in intensity both for the $ZnGa_2O_4$ thin films deposited at low (RT–200°C) and elevated (800°C) temperatures, as illustrated in Fig. 8(c). It is worth noting that a similar emission at 2.48 eV was observed for $ZnGa_2O_4$ films grown by metal organic chemical vapor deposition [54]. The incorporation of Zn into $Ga_2O_3$ to form $ZnGa_2O_4$ would result in a donor–acceptor pair transition, efficiently suppressing the intrinsic green emission band. Furthermore, a recent study of $ZnGa_2O_4$, prepared by a microwave-assisted solid-state reaction, showed green emission centered at 2.35 eV (528 nm), with claims that such electron trapping centers are suitable for persistent luminescence [47]. Thereby, this low energy tail depends significantly on the substrate temperature and the stoichiometric ratio between Ga and Zn. To further enhance the $ZnGa_2O_4$-related optical features, it is suggested to increase cation disorder and zinc/oxygen vacancy concentration [50,55,56], which falls within the scope of our future studies.

Finally, the sharp emission peak in the UV region at 3.29 eV (376 nm; Fig. 8(a)) is attributed to transition levels between the oxygen vacancy and unintended N impurities, introduced during the annealing in nitrogen environment [57]. It should be noted that this sharp feature is manifested in the PL spectra of $ZnGa_2O_4$ thin films grown at lower temperatures (200–400°C) and particularly in the low-temperature PL spectra (typically observed below 200K; see Fig. S7).

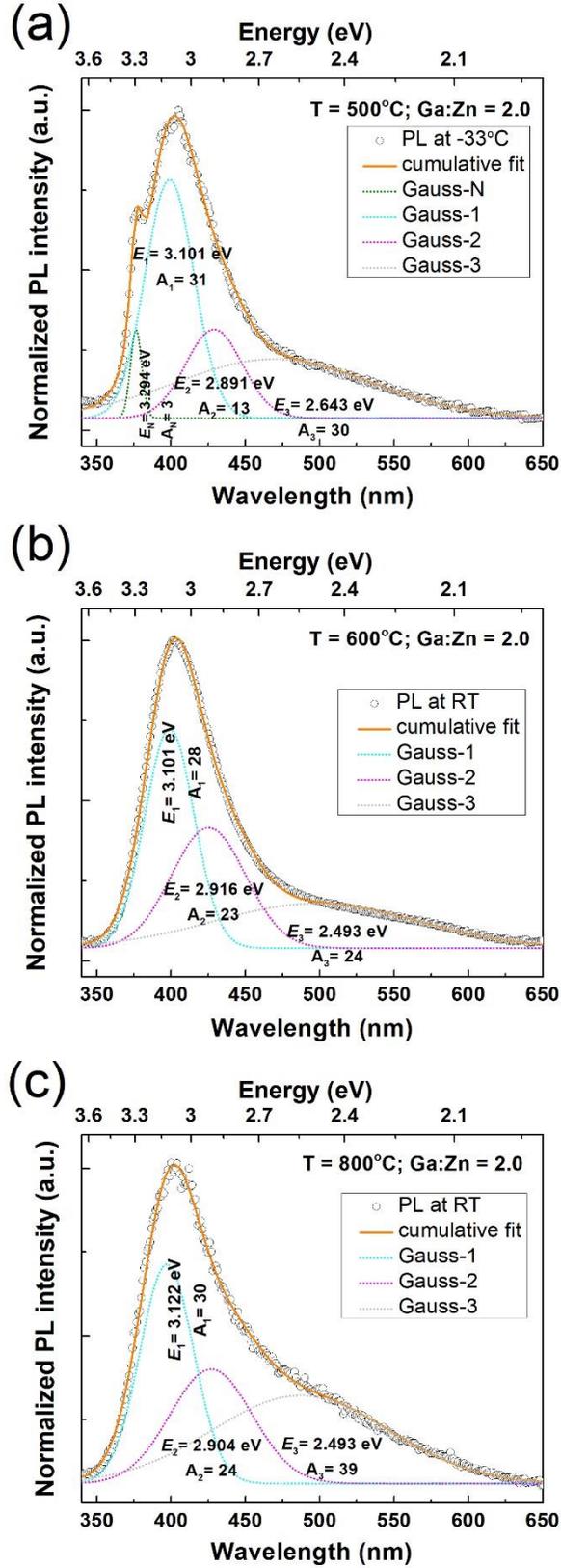

**Fig. 8.** Gaussian component-fit of PL spectra for the ZnGa$_2$O$_4$ thin films deposited at (a) 500°C (PL at -33°C), (b) 600°C (PL at RT), and (c) 800°C (PL at RT) with stoichiometric ratio of Ga:Zn=2.0 with the corresponding central energy *E* and integral area *A* indicated below the corresponding bands.

## Conclusions

We have demonstrated that reactive magnetron co-sputtering is an efficient tool for deposition of ternary oxide films with varied composition even in the relatively complicated case of zinc gallium oxide, where one of the two targets is a liquid, and the Ga:Zn atomic ratio in the film depends strongly not only on the sputtering process parameters, but on the substrate temperature as well. Post-annealing of amorphous films at 700°C in air promoted crystallisation without affecting the composition. Varied sputtering conditions and substrate temperatures yielded amorphous and crystalline films with Ga:Zn ratios ranging from 0.3 to 5.7. The deposition rates were by an order of magnitude higher than those for rf sputtering from ceramic targets. The increase in the Ga:Zn ratio and deposition temperature resulted in the widening of the optical band gap, reaching values of 5.1–5.2 eV, and reduced the absorption in the visible and UVA range initially caused by either the ZnO phase or the amorphous structure of the films.

All the samples deposited in this study were photoluminescent. The spectrum of stoichiometric crystalline $ZnGa_2O_4$ films deposited at 400–600°C substrate temperature consists of a single band centered at 3.1 eV. For the films with compositions other than stoichiometric $ZnGa_2O_4$ and/or substrate temperatures outside the 400–600°C range, the dominant peak at 3.1 eV is accompanied by a broad tail stretching across the green part of the spectrum. The results suggest that the main peak and the low energy tail originate from different types of defects in the film. While the main peak can be ascribed to oxygen-vacancy related transitions, the tail appears to be related to disorder caused by the amorphous structure of the film and/or the excess or deficiency of one of the metal ions.

## Acknowledgements

This study was financially supported via ERDF project No. 1.1.1.1/20/A/057 "Functional ultrawide bandgap gallium oxide and zinc gallate thin films and novel deposition technologies". Photoluminescence measurements by R.N. were supported from post-doctoral research project 1.1.1.2/VIAA/3/19/442. The Institute of Solid State Physics, University of Latvia, as a Center of Excellence, has received funding from the European Union's Horizon 2020 Framework Programme H2020-WIDESPREAD-01-2016-2017-TeamingPhase2 under grant agreement No. 739508, project CAMART².